\documentclass[hessd, manuscript, online]{copernicus}

\usepackage{rotating}
\usepackage{adjustbox}
\usepackage{listings}
\usepackage{ifpdf}

\pdfoutput=1

\begin{document}

\title{Have precipitation extremes and annual totals been increasing in the world's dry regions over the last 60 years?}


\author[1,2]{Sebastian Sippel}

\author[2]{Jakob Zscheischler}
\author[1]{Martin Heimann}
\author[3]{Holger Lange}
\author[1,4,5]{Miguel D. Mahecha}
\author[6]{Geert Jan van Oldenborgh}
\author[7]{Friederike E. L. Otto}
\author[1,4,5]{Markus Reichstein}

\affil[1]{Max Planck Institute for Biogeochemistry, Jena, Germany}
\affil[2]{Institute for Atmospheric and Climate Science, ETH Z{\"u}rich, Z{\"u}rich, Switzerland}
\affil[3]{Norwegian Institute of Bioeconomy Research, \AA s, Norway}
\affil[4]{German Centre for Integrative Biodiversity Research (iDiv), Leipzig, Germany}
\affil[5]{Michael Stifel Center Jena for Data-Driven and Simulation Science, Jena, Germany}
\affil[6]{Weather and Climate Modeling, Koninklijk Nederlands Meteorologisch Instituut, De Bilt, Netherlands}
\affil[7]{Environmental Change Institute, University of Oxford, South Parks Road, Oxford, United Kingdom}


\runningtitle{Precipitation increases in the world's dry regions}

\runningauthor{Sippel et al.}

\correspondence{Sebastian Sippel (ssippel@bgc-jena.mpg.de)}

\received{}
\pubdiscuss{} 
\revised{}
\accepted{}
\published{}


\firstpage{1}

\maketitle

\nolinenumbers

\begin{abstract}
Daily rainfall extremes and annual totals have increased in large parts of the global land area over the last decades. These observations are consistent with theoretical considerations of a warming climate. However, until recently these global tendencies have not been shown to consistently affect land regions with limited moisture availability. A recent study, published by \citet{Donat.2016}, now identified rapid increases in globally aggregated dry region daily extreme rainfall and annual rainfall totals. Here, we reassess the respective analysis and find that 
a) statistical artifacts introduced by the choice of the reference period prior to data standardization lead to an overestimation of the reported trends by up to 40\%, and also that 
b) the definition of `dry regions of the globe' affect the reported globally aggregated trends in extreme rainfall. Using the same observational dataset, but accounting for the statistical artifacts and using alternative, well-established dryness definitions, we find no significant increases in heavy precipitation in the world's dry regions. Adequate data pre-processing approaches and accounting for uncertainties regarding the definition of dryness are crucial to the quantification of spatially aggregated trends in the world's dry regions. In view of the high relevance of the question to many potentially affected stakeholders, we call for a cautionary consideration of specific data processing methods, including issues related to the definition of dry areas, to guarantee robustness of communicated climate change relevant findings. 
\end{abstract}

\clearpage
\introduction  
Daily rainfall extremes are expected to increase roughly by 6-7\% per $^\circ$C of warming following the Clausius-Clapeyron equation \citep{Allen.2002}, if there is enough moisture available. Quantifying and predicting changes in precipitation characteristics due to climate change is crucial for water availability assessments and adaptation to climate change \citep{IPCC.2012,Greve.2014}. On a global scale, daily precipitation extremes have been observed to intensify \citep{Donat.2013,Westra.2013,OGorman.2015}, consistent with global model simulations \citep{Fischer.2015}, and coincide with a global-scale increase in observed annual rainfall totals \citep{Donat.2013}. However, there is little consensus to date on how precipitation characteristics have changed in the past over dry land areas and how they will change in the future. \cite{Donat.2016} investigate whether and to what extent daily precipitation extremes (Rx1d) and annual precipitation totals (PRCPTOT) have increased over the last 60 years using observational data. The authors identify rapid increases in rainfall extremes over dry regions, which strongly outpace the corresponding increases over wet areas, and find a similar pattern for annual rainfall.

The question whether precipitation extremes increase in dry regions is highly relevant in the context of climate change adaptation, as generally dry areas may be less prepared to deal with precipitation extremes \citep{Ingram.2016}. Consequently, the recent report on increasing precipitation extremes in dry areas was highlighted in major Science journals (including \textit{Nature} \citep{Tollefson.2016} and \textit{Nature Climate Change} \citep{Ingram.2016}) and received major media coverage
\footnote{\url{http://www.huffingtonpost.com/entry/global-warming-will-bring-extreme-rain-and-flooding-study-finds_us_56e081c7e4b0860f99d796ab}}$^{,}$
\footnote{\url{https://www.theguardian.com/environment/2016/mar/08/hotter-planet-spells-harder-rains-to-come-study}}$^{,}$
\footnote{\url{
https://www.sciencedaily.com/releases/2016/03/160308105625.htm}}$^{,}$
\footnote{\url{
http://phys.org/news/2016-03-global-world-driest-areas.html}}$^{,}$
\footnote{\url{
http://www.abc.net.au/news/2016-03-08/climate-change-could-bring-more-rain-to-deserts-study/7229236}}$^{,}$
\footnote{\url{
http://www.asce.org/magazine/20160412-climate-change-to-cause-more-precipitation-in-dry-regions,-researchers-say/}}, which indicates the importance of this topic for the scientific community, the public and in the context of climate-related decision-making.

However, scrutinizing the findings by \cite{Donat.2016} reveals two major issues of concern: Firstly, the applied statistical approach introduces two systematic biases that leads to a substantial overestimation of the increase in precipitation totals and extremes of up to 40\% in dry regions. Wet regions, by contrast, are only affected to a limited degree due to a cancellation of errors in trend estimates. Secondly, the definition of dryness used in \cite{Donat.2016} based on precipitation alone does not reflect the common understanding of a dry region and thus induces considerable uncertainty. If we test alternative but well-established definitions of a `dry region' \cite[based on water supply and demand, either implicitly or explicitly, cf.][]{Koppen.1900,Greve.2014} and reapply the appropriate statistical tools, we find no significant increases in precipitation extremes or annual totals in the world's dry regions. The increases in wet regions are slightly underestimated by \cite{Donat.2016} due to a cancellation of errors. These results are of high relevance in the context of making climate change adaptation decisions in dry regions.

\section{On data pre-processing based on a time-invariant reference period}
As a first step in the analysis of \cite{Donat.2016}, the authors normalize the 60-year time series in the gridded HadEX2 dataset \citep{Donat.2013} for each grid point with the sample mean of a 30-year reference period (1951--1980), which is a widespread procedure in climate science \citep{Zhang.2005,Sippel.2015}. However, this procedure artificially increases the mean and spread of the spatial distribution in the out-of-base period (1981-2010) in all investigated time series, simply because variability in the sample means inflates the signal in the latter period \citep{Sippel.2015} (Fig.~\ref{fig1}). To illustrate this point, consider two hypothetical climate regions of the same size: In region one, the mean of a rainfall quantity increases between two periods (from 100 to 200mm, say), for example due to a few large extremes, whereas it decreases by exactly the same amount in region two (i.e. from 200 to 100mm). Consequently, in both time periods the spatial average and the spread of the two regions would be statistically indistinguishable. However, normalizing by the mean of the first time period would imply that the spatial average across both regions for the second period is 1.25 (the average of 0.5 and 2), i.e. a spurious increase of 25\% between both periods. In addition, the normalization induces considerable spread in the spatial distribution in the second period.

An additional statistical bias stems from the choice of the world's 30\% wettest and 30\% driest regions based on the climatology of PRCPTOT and Rx1d in the reference period (1951-1980). Because 30 years are fairly short to derive a robust climatology of the tails of the precipitation distribution, the computed changes in wet and dry regions are distorted by the "regression to the mean" phenomenon \citep{Galton.1886,Barnett.2005}. In other words, selecting from the dry (wet) end of the spatial distribution in one subset of the dataset will result in a higher probability for wetter (drier) conditions in the remaining years (Figure in Appendix A, Tables in Appendix B).

Hence, the chosen normalization approach combined with the spatial point selection method gives a bias toward precipitation totals and extremes  increasing at a faster rate in dry regions compared to wet regions. Yet over dry regions, these two effects lead to an overestimation of the trends in rainfall totals and extremes by 40.3\% and 33.2\%, respectively (Fig.~\ref{fig2}, Tables in Appendix B, reference: 1951-2010). In contrast, in wet regions these effects roughly cancel each other out in the case of extremes (+8.7\%) and lead to an underestimation of the increase in total rainfall (-13.7\%).

\section{On the definition of a dry region}
Climatological dryness is generally not determined by water supply alone as implied by the definition in \cite{Donat.2016} but also depends on atmospheric water demand, that is the ability to evaporate water from the land surface \citep{Koppen.1900}. This means that "\textit{we cannot tell whether a climate is moist or dry by knowing precipitation alone; we must know whether precipitation is greater or less than potential evapotranspiration}", as Charles Warren Thornthwaite put it in a landmark paper \citep{Thornthwaite.1948}; a statement that is clearly mirrored in present-day literature \citep[e.g.][]{Greve.2014}. For the analysis of precipitation extremes (Rx1d), regions in northern Europe such as parts of Scandinavia or the Netherlands fall in the `dry' class because of relatively small annual maximum daily rainfall (Figure in Appendix~A and Fig. 1c in \citet{Donat.2016}). This is in contrast to what is commonly understood by the term `dry' and consequently induces considerable confusion regarding increases in rainfall extremes and totals in globally `dry' regions. 
To clarify this issue, we also test the sensitivity of the reported increases in Rx1d to the choice of dryness definition by using a variety of different dryness definitions (Figure in Appendix~A). Hence, we evaluate trends and period increments in Rx1d and PRCPTOT in 
\begin{enumerate}
\item regions that fall below the global 30\% quantile in HadEX2 in the respective diagnostic (Rx1d or PRCPTOT), following \citet{Donat.2016},
\item dry regions (`B-climates') from a traditional climate classification based on temperature and precipitation  \citep{Koppen.1900,Kottek.2006}, 
\item dry regions as identified from an aridity-based definition of dryness \citep{Greve.2014}, and
\item dry and transitional regions combined from the latter definition \citep{Greve.2014}. 
\end{enumerate}
In addition, we test uncertainties related to the temporal coverage of the dataset by relying on time series with 90\% coverage (cf. \citet{Donat.2016}) and in addition we analyse only time series without missing values (100\% coverage). 

Our results show that, based on a two-sided trend test, no significant increases in precipitation totals and extremes can be detected in the world's dry regions as defined based on either \citet{Greve.2014} or \citet{Koppen.1900}, or in dry and transitional regions combined or if time series with incomplete temporal coverage are removed (Fig.~\ref{fig3}, Table 1-2). These results are consistent with earlier studies that report no or modest changes in Rx1d and PRCPTOT in (predominantly dry) subsidence regions based on model simulations \citep{Kharin.2007,Fischer.2015} and in observations for individual subtropical regions such as Australia or the Mediterranean \citep{Westra.2013,Lehmann.2015}. In contrast, if `the world's dry regions' are defined based on falling below a global 30\% threshold in Rx1d or PRCPTOT in the HadEX2 dataset \citep{Donat.2016}, we indeed confirm robust increases in both Rx1d and PRCPTOT. Thus, the reported increases in both diagnostics are highly sensitive to the definition of a `dry region', and appear to stem from regions with relatively moderate extreme (Rx1d) or average (PRCPTOT) rainfall, such as regions in Northern Europe (Rx1d, Figure in Appendix~A) or North-East Siberia (PRCPTOT, Figure in Appendix~A) that are not commonly understood as `dry'.
Hence, we conclude extreme rainfall (Rx1d) or annual totals (PRCPTOT) do not show significant increases in the world's dry (i.e. arid, or arid-transitional) regions in conventional definitions.

\conclusions  
An accurate quantification of spatially aggregated trends in a rapidly changing Earth system is of highest relevance to decision-makers in the context of climate change adaptation \citep{Fischer.2015}. Therefore, short reference periods that are defined on a subset of the available dataset for normalization or data pre-processing purposes \citep{Zhang.2005,Sippel.2015} should be avoided, as this procedure inevitably introduces biases. Furthermore, the definition of a `dry region' induces considerable uncertainty in quantifying changes in rainfall extremes or totals. We find no evidence for changes in observed annual rainfall or heavy precipitation in the world's dry or dry-transitional regions if the notion of dryness is based on water supply and demand (i.e. aridity) and biases due to short reference periods are avoided. Thus, understanding and disentangling the discrepancy in rainfall trends between regions of moderate extreme or total rainfall \citep{Donat.2016} and data-scarce arid regions, and relating both to model simulations, remains a research priority of high relevance.



\clearpage
\authorcontribution{S.S. and J.Z. conceived the study. All authors contributed to writing the paper.}

\begin{acknowledgements}
S.S. and M.D.M. are grateful to the European Commission for funding the BACI project (grant agreement No 640176) and to the European Space agency for funding the STSE project CAB-LAB.
\end{acknowledgements}

\clearpage
\appendix
\section{}
\begin{figure}[h!]
\centering
\includegraphics[width=0.9\textwidth]{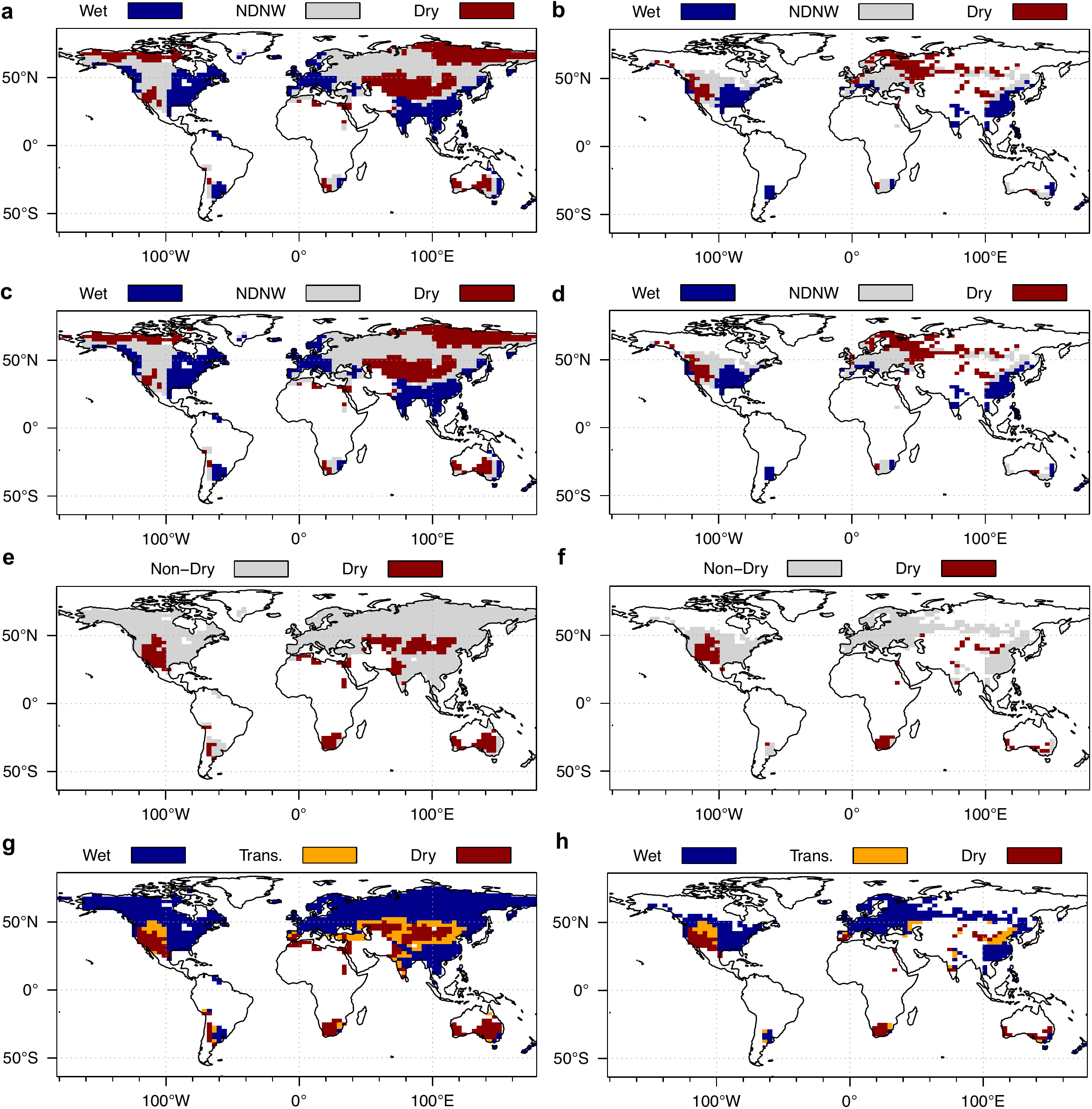}
\caption*{Mask of the world's dry and wet region: Avoiding `regression to the mean'.
a-d, Dryness masks based on 1951-1980 and HadEX2 (a-b, see Donat et al. (2016)) and 1951-2010 (c-d, to avoid `regression to the mean' selection bias, see main text) for PRCPTOT (left) and Rx1d (right). `NDNW' indicates neither dry nor wet areas, white inland areas indicate less than 90\% data availability in the HadEX2 dataset and were not considered.
e-f, Dry regions based on Koeppen (1900) as updated by Kottek et al. (2006) and intersected by data availability in HadEX2.
g-h, Dry and transitional regions following Greve et al. (2014) and intersected by data availability in HadEX2. 
}
\label{SI_fig1}
\end{figure}

\clearpage
\section{}
\scriptsize

\begin{figure}[h!]
\centering
\includegraphics[width=0.98\textwidth]{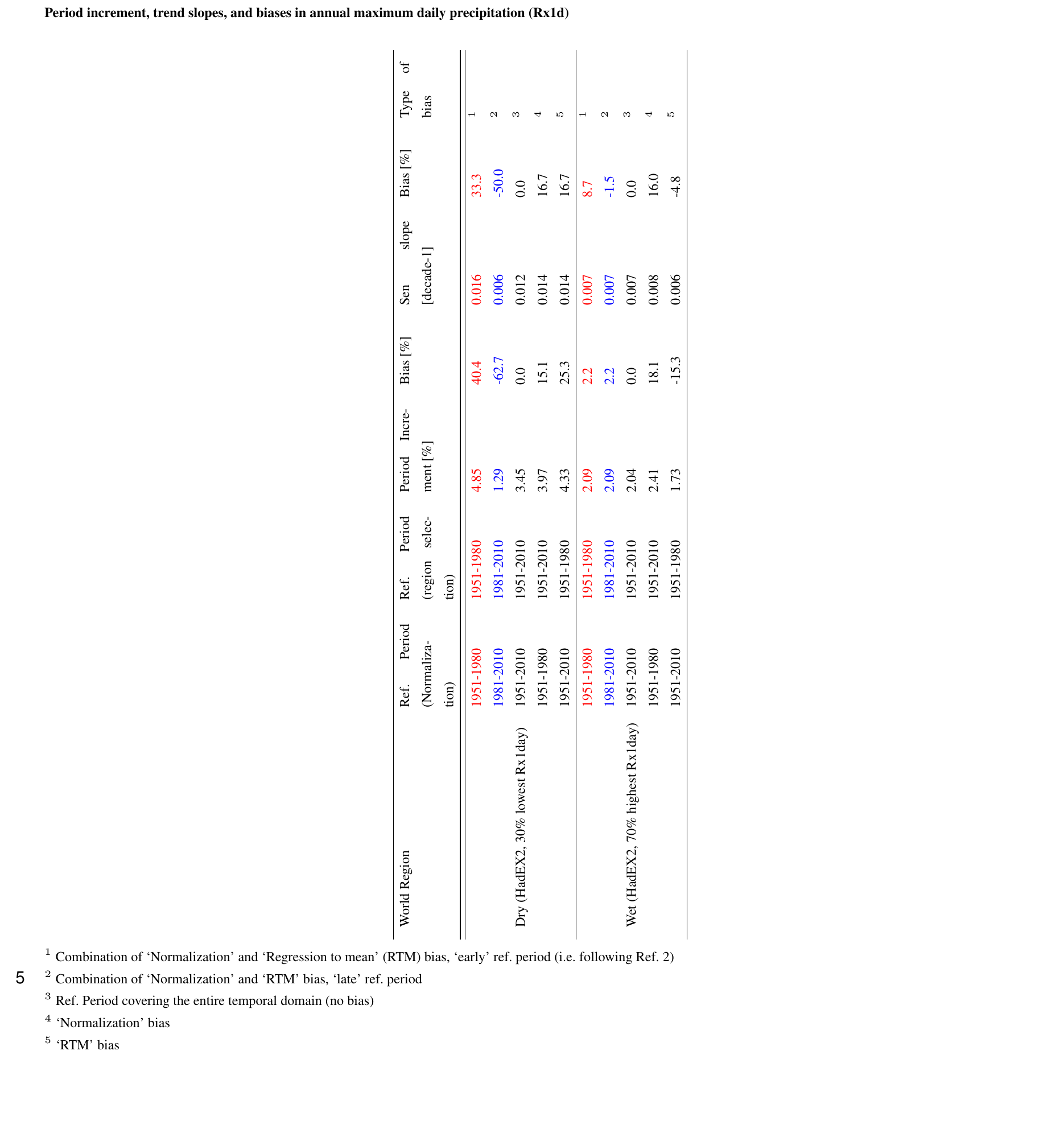}
\end{figure}

\begin{figure}[h!]
\centering
\includegraphics[width=0.98\textwidth]{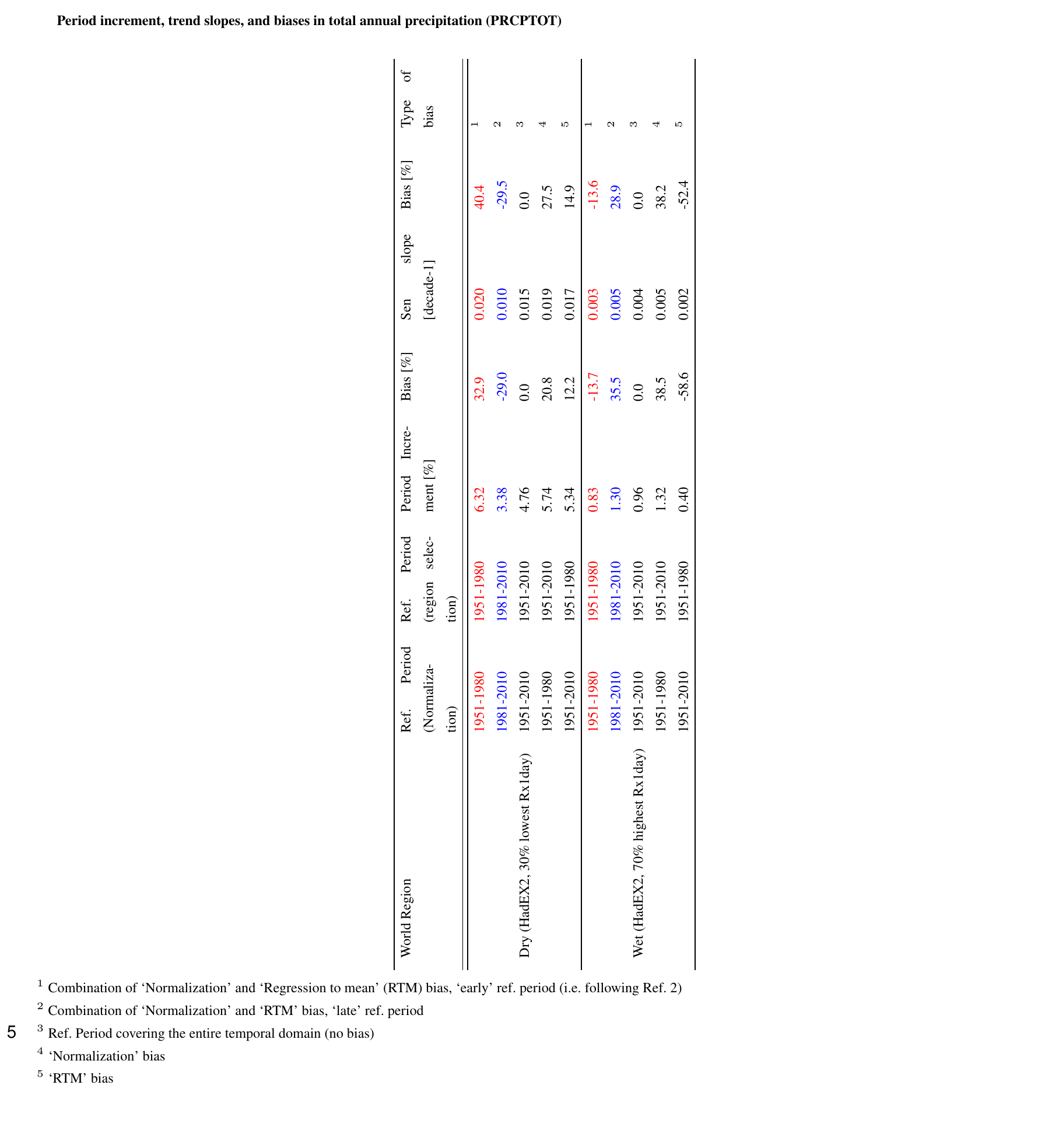}
\end{figure}







\bibliographystyle{copernicus}
\bibliography{normalization.bib}




%
%

\clearpage

\begin{figure*}[t]
\includegraphics[width=1\textwidth]{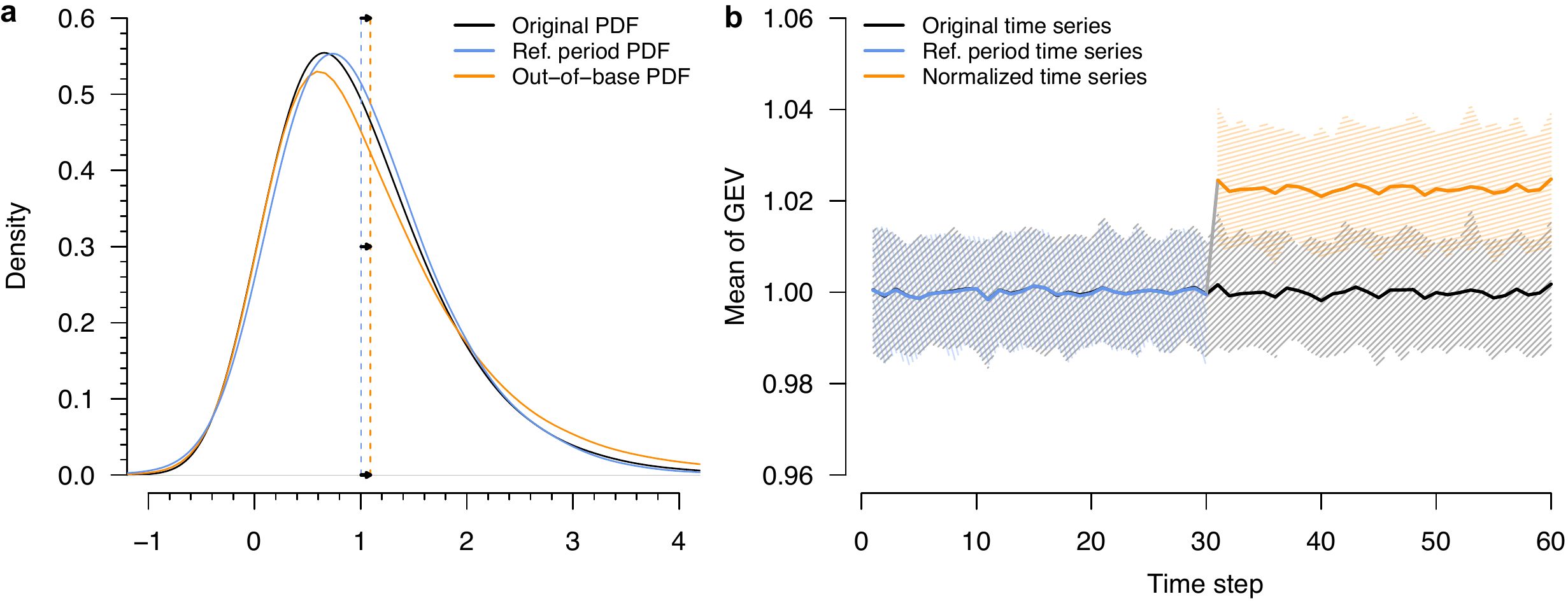}
\caption{Conceptual example of biases in the mean induced by normalization based on a fixed reference period.
a, Probability distributions and their respective means for an artificial dataset of $10^4$ grid cells each comprised of  random variables sampled from a Generalized Extreme Value distribution (GEV, $\mu = 1$, $\sigma = 1$, $\xi = 0$, sample size $n_{ref}=8$ for illustration), and normalized following \citet{Donat.2016} with different ref. periods. b, Shift in the mean of spatially aggregated variables due to reference period normalization ($n_{ref}=30$ following \citet{Donat.2016}, Confidence intervals denote the 5th - 95th percentile). Code to reproduce this example is provided in Supplementary Material.}
\label{fig1}
\end{figure*}

\begin{figure*}[t]
\includegraphics[width=1\textwidth]{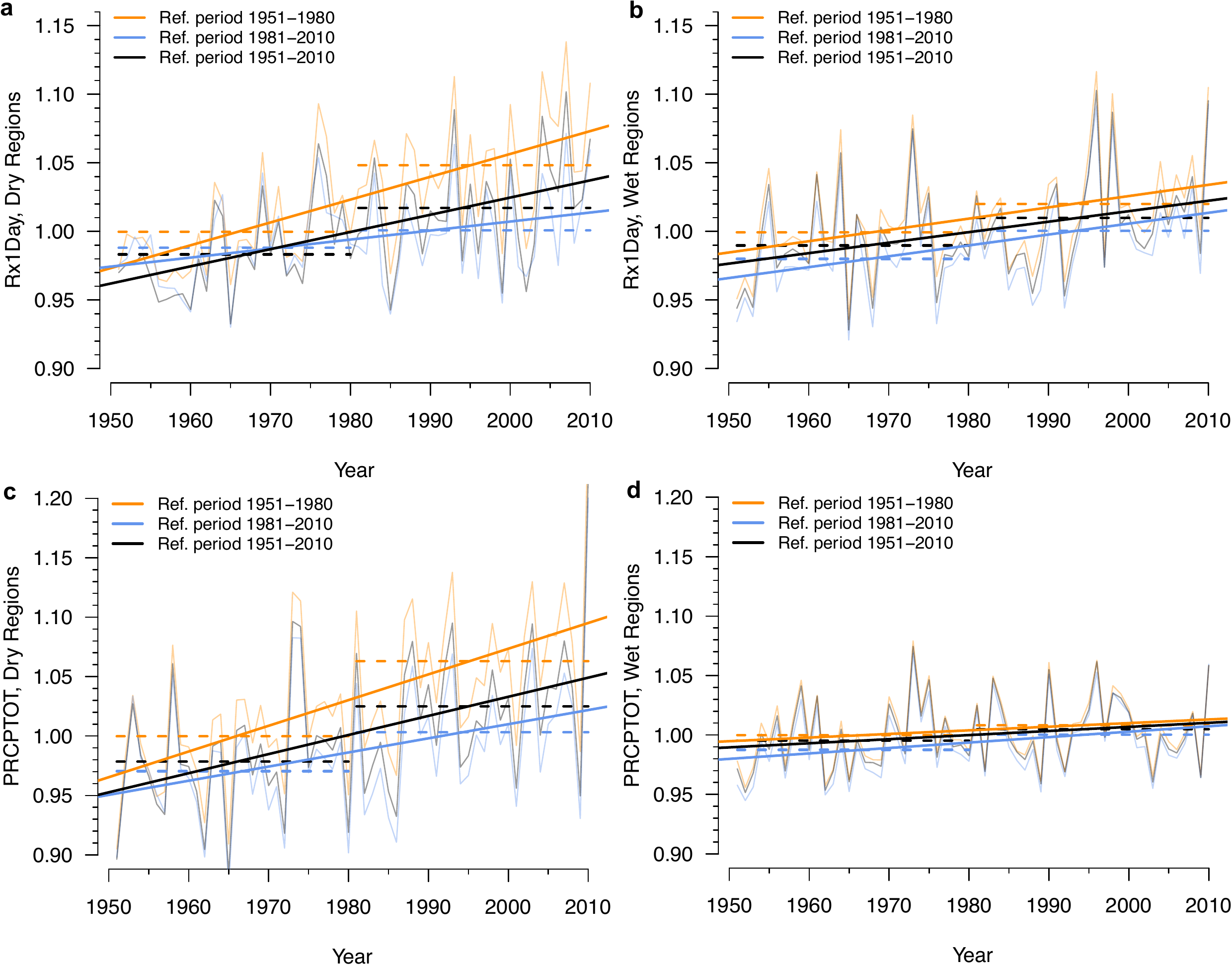}
\caption{Normalization-induced biases on time series and trend estimates.
a-b, Time series, trends and 30-year means of spatially aggregated heavy precipitation (Rx1d) in (a) dry  and (b) wet regions. c-d, Time series, trends and 30-year means of spatially aggregated total precipitation (PRCPTOT) in (a) dry  and (b) wet regions. Red lines are taken from \citet{Donat.2016} (ref. period: 1951-1980), black lines are corrected for biases (ref. period: 1951-2010), and blue lines indicate a hypothetical 1981-2010 reference period.
}
\label{fig2}
\end{figure*}

\begin{figure*}[t]
\includegraphics[height=0.85\textheight]{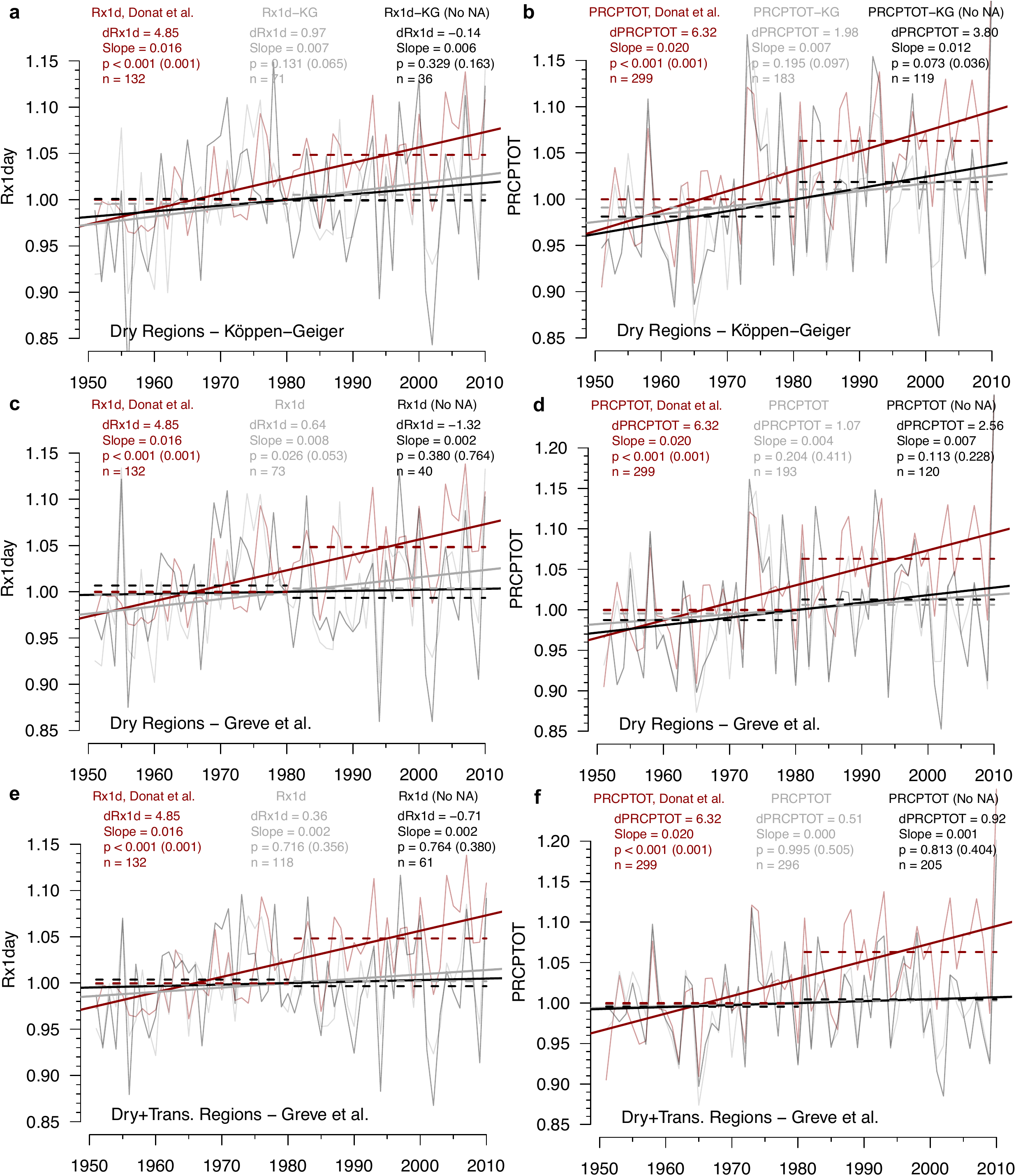}
\caption{a-f, Time series, trends and 30-year means of spatially aggregated heavy precipitation (Rx1d, a,c,e) and annual rainfall totals (PRCPTOT, b,d,f) in dry regions following (a-b) \citet{Koppen.1900}, (c-d) \citet{Greve.2014}, and dry and transitional regions combined \citep{Greve.2014}. a-b, Red lines as reported in \citet{Donat.2016} for comparison, i.e. based on the 1951-1980 reference period. Grey lines indicate our estimation using an aridity-based dryness definition and correcting for statistical artefacts, black lines are estimated equally, but after removing cases of incomplete temporal coverage. All p-values are given for two-sided (one-sided) Mann-Kendall trend tests. 
}
\label{fig3}
\end{figure*}

%
%
%
%
%
%
%
%
%
%
%
%
\begin{table*}[t]
\caption{Uncertainties regarding the definition of a `dry region', Rx1d.}
\centering
\begin{tabular}{p{7cm}p{1.5cm}p{1.5cm}p{1.5cm}p{1.5cm}p{1.3cm}p{1.2cm}}
 \tophline
Dry Region Definition & Ref. Period & Temporal Coverage (\%) & Period Inc. (\%) & Trend Slope (\% per year) & two-sided p-value (one-sided) & Sample size \\
 \middlehline
\citet{Donat.2016}, global 30\% quantile in Rx1d & 1951-1980 & 90\% & $4.85$ & $0.016$ & $<0.001$ ($<0.001$) & 132 \\
\citet{Donat.2016}, global 30\% quantile in Rx1d & 1951-1980 & 100\% & $6.07$ & $0.020$ & $<0.001$ ($<0.001$) & 57 \\
\citet{Donat.2016}, global 30\% quantile in Rx1d & 1951-2010 & 90\% & $3.45$ & $0.012$ & $<0.001$ ($<0.001$) & 132 \\
\citet{Donat.2016}, global 30\% quantile in Rx1d & 1951-2010 & 100\% & $4.24$ & $0.017$ & $<0.001$ ($<0.001$) & 57 \\
\citet{Koppen.1900}, dry climates (`B-climates') & 1951-2010 & 90\% & $0.97$ & $0.007$ & $0.131$ ($0.064$) & 71 \\
\citet{Koppen.1900}, dry climates (`B-climates') & 1951-2010 & 100\% & $-0.14$ & $0.006$ & $0.329$ ($0.163$) & 36 \\
\citet{Greve.2014}, dry regions & 1951-2010 & 90\% & $0.64$ & $0.008$ & $0.053$ ($0.026$) & 73 \\
\citet{Greve.2014}, dry regions & 1951-2010 & 100\% & $-1.32$ & $0.002$ & $0.764$ ($0.380$) & 40 \\
\citet{Greve.2014}, dry+transitional regions & 1951-2010 & 90\% & $0.36$ & $0.005$ & $0.195$ ($0.097$) & 118 \\
\citet{Greve.2014}, dry+transitional regions & 1951-2010 & 100\% & $-0.71$ & $0.002$ & $0.716$ ($0.356$) & 61 \\
  \bottomhline
\end{tabular}
\end{table*}

\begin{table*}[t]
\caption{Uncertainties regarding the definition of a `dry region', PRCPTOT.}
\centering
\begin{tabular}{p{7cm}p{1.5cm}p{1.5cm}p{1.5cm}p{1.5cm}p{1.3cm}p{1.2cm}}
 \tophline
Dry Region Definition & Ref. Period & Temporal Coverage (\%) & Period Inc. (\%) & Trend Slope (\% per year) & two-sided p-value (one-sided) & Sample size \\
 \middlehline
\citet{Donat.2016}, global 30\% quantile in PRCPTOT & 1951-1980 & 90\% & $6.32$ & $0.020$ & $<0.001$ ($<0.001$) & 299 \\
\citet{Donat.2016}, global 30\% quantile in PRCPTOT & 1951-1980 & 100\% & $5.93$ & $0.015$ & $0.002$ ($0.001$) & 108 \\
\citet{Donat.2016}, global 30\% quantile in PRCPTOT & 1951-2010 & 90\% & $4.76$ & $0.015$ & $<0.001$ ($<0.001$) & 299 \\
\citet{Donat.2016}, global 30\% quantile in PRCPTOT & 1951-2010 & 100\% & $4.37$ & $0.010$ & $0.157$ ($0.077$) & 108 \\
\citet{Koppen.1900}, dry climates (`B-climates') & 1951-2010 & 90\% & $1.98$ & $0.007$ & $0.195$ ($0.100$) & 183 \\
\citet{Koppen.1900}, dry climates (`B-climates') & 1951-2010 & 100\% & $3.80$ & $0.012$ & $0.073$ ($0.036$) & 119 \\
\citet{Greve.2014}, dry regions & 1951-2010 & 90\% & $1.07$ & $0.004$ & $0.411$ ($0.204$) & 193 \\
\citet{Greve.2014}, dry regions & 1951-2010 & 100\% & $2.56$ & $0.007$ & $0.228$ ($0.107$) & 120 \\
\citet{Greve.2014}, dry+transitional regions & 1951-2010 & 90\% & $0.51$ & $<0.000$ & $0.995$ ($0.505$) & 296 \\
\citet{Greve.2014}, dry+transitional regions & 1951-2010 & 100\% & $0.92$ & $0.001$ & $0.813$ ($0.404$) & 205 \\
  \bottomhline
\end{tabular}
\end{table*}

\end{document}